# MICADO: The Multi-Adaptive Optics Camera for Deep Observations

Richard Davies[1]
Veronika Hörmann[1]
Sebastian Rabien[1]
Eckhard Sturm[1]
João Alves[2]
Yann Clénet[3]
Jari Kotilainen[4]
Florian Lang-Bardl[5]
Harald Nicklas[6]
Jörg-Uwe Pott[7]
Eline Tolstoy[8]
Benedetta Vulcani[9]
and the MICADO Consortium[a]

[1] Max Planck Institute for Extraterrestrial Physics, Garching, Germany
[2] University of Vienna, Austria
[3] LESIA, Université PSL, CNRS, Sorbonne Université, Université de Paris, Observatoire de Paris, France
[4] FINCA, University of Turku, Finland
[5] Universitäts-Sternwarte, Munich, Germany
[6] Georg-August-Universität, Göttingen, Germany
[7] Max Planck Institute for Astronomy, Heidelberg, Germany
[8] Kapteyn Institute, Groningen, the Netherlands
[9] INAF – Osservatorio di Padova, Italy

The Multi-adaptive optics Imaging CamerA for Deep Observations (MICADO) will image a field of view of nearly 1 arcminute at the diffraction limit of the Extremely Large Telescope (ELT), making use of the adaptive optics correction provided by single-conjugate adaptive optics (SCAO) and multi-conjugate adaptive optics (MCAO). Its simple and robust design will yield an unprecedented combination of sensitivity and resolution across the field. This article outlines the characteristics of the observing modes offered and illustrates each of them with an astrophysical application. Potential users can explore their own ideas using the data simulator ScopeSim.

## Introduction

MICADO will provide the ELT with a diffraction-limited capability for imaging, coronagraphy, and slit spectroscopy at near-infrared wavelengths. The instrument is optimised to work with the laser-guide-star MCAO system developed by the Multi-conjugate Adaptive Optics RelaY (MAORY) consortium. It will also have a SCAO mode that uses just a single natural guide star. Following the start of Phase B in October 2015, MICADO had its Preliminary Design Review in November 2018, and is ready for its Final Design Review in 2021. The current plan is that, after an initial phase of operations at the ELT's first light, during which MICADO will be available only with the SCAO system, it will move to its final configuration where it interfaces to MAORY and will benefit from both a SCAO and a MCAO correction.

MICADO will be able to address many science topics relevant to modern astrophysics, and has clear synergies with other instruments and facilities. The science cases that have driven the design focus on five main themes: (i) galaxy evolution at high redshift, (ii) black holes in galaxy centres, including the centre of the Milky Way, (iii) resolved stellar populations, including photometry in galaxy nuclei, the initial mass function in young star clusters, and intermediate-mass black holes in globular clusters, (iv) characterisation of exoplanets and circumnuclear discs at small angular scales, and (v) the Solar System. To address these, MICADO will exploit its sensitivity and resolution in four observing modes: standard imaging, astrometric imaging, coronagraphic imaging, and spectroscopy. Both SCAO and MCAO can be used with all observing modes, albeit with some limitations, and the choice depends on the specific scientific goals as well as the target itself. Details about the MCAO system, and some further science applications, are given in the accompanying article on MAORY (Ciliegi et al., p. 13).

## Observing modes

### Standard imaging

Standard imaging is the simplest observing mode, designed to obtain images at the diffraction-limited resolution of 4–12 milliarcseconds at wavelengths of

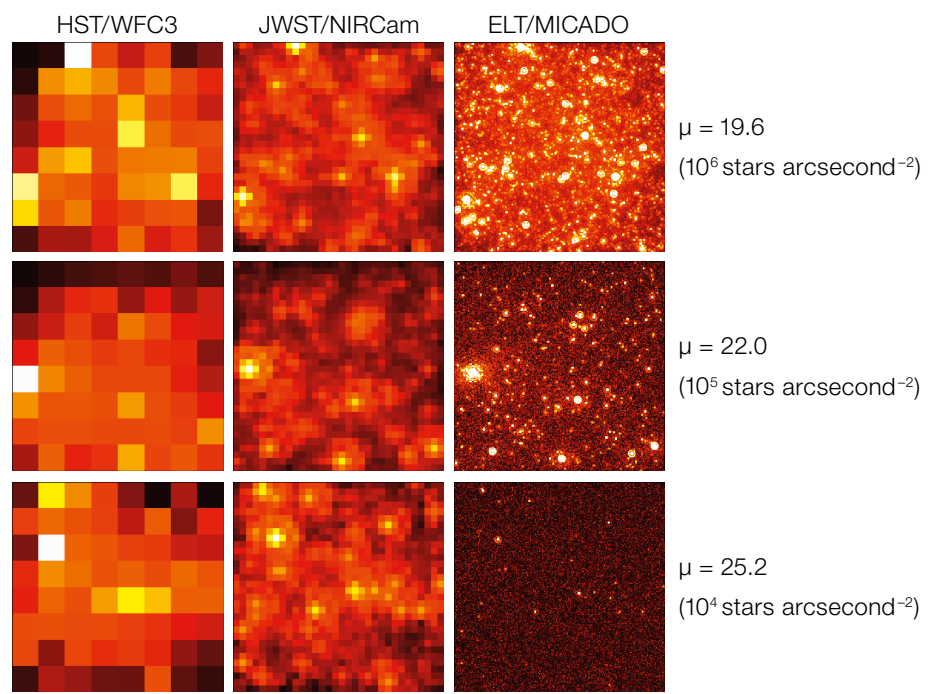

Figure 1. Comparison of how crowded stellar fields would appear when observed by the HST (left), the JWST (centre), and MICADO (right). The bottom row matches the stellar density at a radius of 4–5 $R_{eff}$ for NGC 4472 in the Virgo Cluster and represents the limit of JWST resolution. The top row corresponds to 2 $R_{eff}$ in the same galaxy and many individual stars can still be measured by MICADO. Each panel is 1 arcsecond across. These simulations were performed with ScopeSim.

HST/WFC3　　JWST/NIRCam　　ELT/MICADO

µ = 19.6
($10^6$ stars arcsecond$^{-2}$)

µ = 22.0
($10^5$ stars arcsecond$^{-2}$)

µ = 25.2
($10^4$ stars arcsecond$^{-2}$)

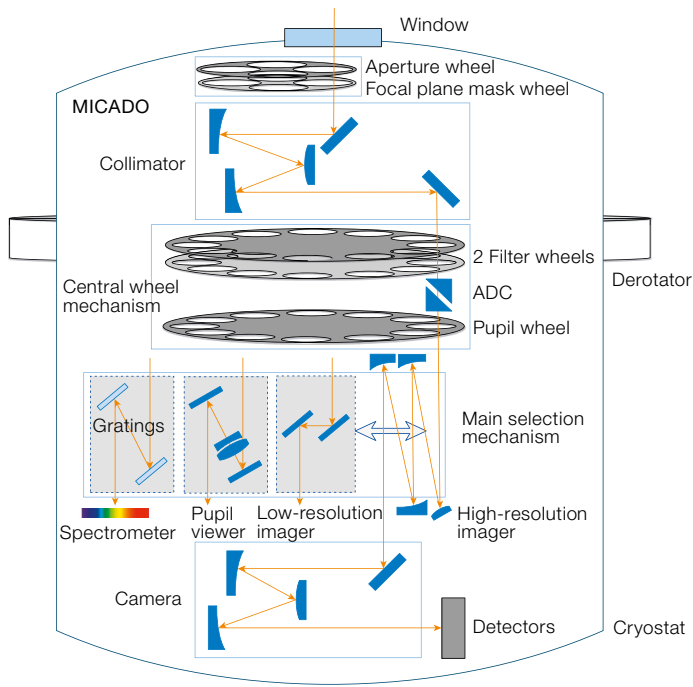

Figure 2. Schematic view of the MICADO design concept, illustrating how the cold optics and mechanisms are assigned to separate modules that can be tested separately and then integrated together in the cryostat.

0.8–2.4 μm. It can be used with SCAO, but will benefit enormously from the MCAO capability, which provides a uniform point spread function (PSF) over the field as well as high sky coverage. An array of 3 × 3 H4RG detectors (150 million pixels) provides a field of view of 50.5 × 50.5 arcseconds at a pixel scale of 4 milliarcseconds with the low-resolution imager. Complementing the multiplex advantage of this configuration is the high-resolution imager with a 1.5-milliarcsecond pixel scale over a 19 × 19 arcsecond field of view. This fully samples the diffraction-limited PSF from 0.8 μm to 1.5 μm, and provides the fine sampling at the longer wavelengths needed for PSF de-blending in very crowded fields. An atmospheric dispersion corrector ensures that the PSF remains compact (indeed, the chromatic dispersion is large enough that, even for isolated point sources, the gain in sensitivity over most of the sky outweighs the loss in throughput due to the additional optics). And, while in typical observatories more than 90% of observations are performed with no more than 10 filters, MICADO offers considerable flexibility with its large filter wheels which are able to hold more than 30 filters.

With a point-source sensitivity comparable to that of the James Webb Space Telescope (JWST) and a resolution about a factor of 6 better, this mode is well suited to numerous science cases. For the specific topic of galaxy evolution over cosmic time, we now have a fairly robust outline, in terms of global properties, of how galaxies assembled and transformed into the present-day Hubble sequence. The next step is to resolve faint distant galaxies on sufficiently small scales, to assess their sub-galactic components (disc structures, nascent bulges, clumps, and globular cluster progenitors) at spatial scales < 100 pc — equivalent to the seeing limit for the nearby Virgo Cluster galaxies. Relatively unexplored regimes include lower-mass galaxies, comprising the bulk (by number) of the galaxy population, and galaxies at early cosmic times when they were building their first stars.

An alternative probe of galaxy evolution is via relic populations in local galaxies, performing photometry on individual stars to generate colour-magnitude diagrams. Detecting stars on the horizontal branch enables one to trace the star formation history of local galaxies to the reionisation epoch at redshift $z > 6$. The ultimate goal for resolved stellar populations is to measure individual stars in the central regions of elliptical galaxies in the Virgo Cluster. As can be seen in Figure 1, the extreme stellar crowding makes this very challenging. The JWST will be most effective in the outskirts of these galaxies, while the higher resolution of MICADO will enable it to probe within the central effective radius.

Astrometric imaging

One of the most challenging requirements for MICADO is to perform astrometry over the full field at a precision better than 50 microarcseconds, with a goal of 10 microarcseconds — which is comparable to that achieved by the GRAVITY instrument at ESO's Very Large Telescope Interferometer (VLTI) and dedicated space missions such as Gaia, but with much fainter stars. To understand what that requirement means, we must distinguish between absolute and relative astrometry. Absolute astrometry is needed when comparing the positions of objects observed with different instruments, often in different wavebands. Because it relies on an external reference frame, which is dependent on the target field, it can only be done on a best effort basis. On the other hand, in the case of the MICADO instrument, the requirement refers to relative (or differential) astrometry, which is about changes in position between epochs and focuses on proper motion rather than position.

In this context, an obvious astrophysical rationale is the use of stellar proper motions to probe the existence and masses of black holes in stellar clusters and nearby low-mass dwarf galaxies. Studies of globular clusters have yielded a variety of tantalising results, but without robust detections. One key question concerns the relation between the mass of the central black hole and the velocity dispersion of the stellar spheroid around it. Intriguingly, a compilation of black hole mass limits for globular clusters concluded that the slope of the relation for those is rather shallower than that for galaxy bulges and elliptical galaxies. This implies a different regulation process in the star clusters, perhaps suggesting that many of those systems may be the stripped nuclei of dwarf galaxies. Again, current facilities are limited by the extreme crowding that occurs in the centres of the star clusters, exactly where the key measurements have to be made. In particular, proper motions, rather than just line-of-sight velocity dispersions, are essential to measure the anisotropy, which can have a significant impact on the black hole mass derived. Suitable measurements that overcome the crowding will only become possible with the spatial resolution of the ELT.

Meeting this requirement is a major challenge (Pott et al., 2018). To do so, stability and calibration are more important than solely minimising the geometric distortion, which is < 0.4% and < 1.2% for the low- and high-resolution imagers, respectively. And it is necessary to distinguish between linear distortions over the full field, low-order distortions that affect ~ 10-arcsecond scales, and high-order distortions which dominate at < 1-arcsecond scales. These are handled in different ways most appropriate for each case, whether via instrument design (minimising mechanical and thermal flexure), operational scenario (use of calibration masks, and if needed, observational constraints), or post-processing (correcting low-order drifts between individual frames). The relative uniformity of the PSF enabled by MCAO is clearly an important factor in achieving the most precise astrometry.

Coronagraphic imaging

The study of planets around other stars is one of the fundamental science drivers for the ELT. Now that a large number of exoplanets are known, we are entering a phase driven by the need to characterise these planets, in particular the atmospheres of giant exoplanets. Direct imaging of exoplanets provides an opportunity to do this through the use of intermediate band filters that cover molecular absorption bands, enabling one to distinguish models with different temperatures, surface gravities, and clouds. Comparing observations with the Spectro-Polarimetric High-contrast Exoplanet REsearch instrument (SPHERE) to MICADO simulations for the planetary system around HR 8799 provides a glimpse of what it may be possible to achieve. Even in raw images, the inner two known planets are visible. And after basic processing it becomes possible to identify and characterise simulated fainter planets closer in. It opens up the very exciting potential to image planets for which a mass estimation is available from Gaia.

By exploiting the large aperture of the ELT, MICADO will achieve a meaningful contrast at very small inner working angles, and hence serve also as a pathfinder for future dedicated instrumentation. It will improve on a similar system on the VLT in a number of aspects: probing a factor of 5 closer to the primary star, increasing by a factor of 25 the contrast between the exoplanet PSF and the speckle noise that dominates in these locations, and elongation of the speckles which provides better discrimination between those and a compact exoplanet. The focus for MICADO will therefore be in terms of exoplanets at small orbital separations (~ 1 au) around nearby stars (< 20 pc), exoplanets at larger separations (> 10 au) around more distant stars (> 100 pc), and the circumstellar discs from which they form.

High-contrast imaging is an important driver for the SCAO system. The coronagraphy itself can make use of pupil tracking to enable angular differential imaging, and will be achieved via focal-plane and pupil-plane phase masks (in the former case, the centring of the star is actively maintained during the observation). Only basic post-processing is provided since algorithms evolve fast and the best one to use depends on the specific application. MICADO includes a pupil imager to ensure there is a detailed record of how the pupil appears, since the ELT pupil can change between nights, owing to "missing" segments and because the maintenance schedule leads to changing individual segment transmissions.

Spectroscopy

The rationale for spectroscopy in MICADO is to emulate the success of X-shooter while addressing a role complementary to the integral field spectroscopic capability of the High Angular Resolution Monolithic Optical and Near-infrared Integral field spectrograph (HARMONI, Thatte et al., p. 7). With a focus on faint compact or unresolved objects, MICADO uses a slit and provides a resolution of $R \sim 20\,000$, corresponding to 15 km s$^{-1}$ in the $H$ band. The full bandpass is covered in just two settings: a short slit for 0.84–1.48 μm, and a long slit for 1.48–2.46 μm (and also 1.16–1.35 μm) that enables extraction of the sky background further off-axis. This is realised with a fixed configuration cross-dispersing spectrograph module. The wavelength range is defined solely by choice of slit and order-sorting filter. Since the focus is on compact objects, the role of MCAO here is to maximise the Strehl ratio in the centre of the field with a higher sky coverage than can be reached with SCAO.

There are a wide range of applications for this capability. At high redshift, one can measure the emission-line spectra of early supernovae or continuum absorption features in early-type galaxies to measure stellar populations and dynamics. More locally, one can use the line-of-sight velocity dispersions of nearby galaxies to constrain orbit-based models and hence derive black hole masses in galaxy nuclei, extending the currently accessible parameter space to lower black hole masses as well as more distant galaxies. In the Galactic centre, an exciting opportunity is to measure the spin of the black hole, a goal that is more tractable via spectroscopy than astrometry. This can be achieved by tracking a late-type star whose entire orbit lies within ~ 10 milliarcseconds (0.5 light days, about 1/10 of the S2 orbit) so that it is spatially indistinguishable from Sgr A* itself. Spectrally, determining the relative velocity of the star to a precision of < 1 km s$^{-1}$ enables one to discern the impact on its orbit of the black hole's quadrupole moment, which according to general relativity is fully determined by the spin. Even though in practice it is difficult to completely stabilise a source in the slit, sufficient precision can be reached via internal referencing between the stellar absorption features and the atmospheric absorption features imprinted into the observed continuum. Lastly, in terms of exoplanets, MICADO offers the potential for obtaining detailed spectra — with applications in both chemical and dynamical analyses — simply by exploiting the enhanced contrast due to the higher angular resolution of the ELT, without the need for simultaneous

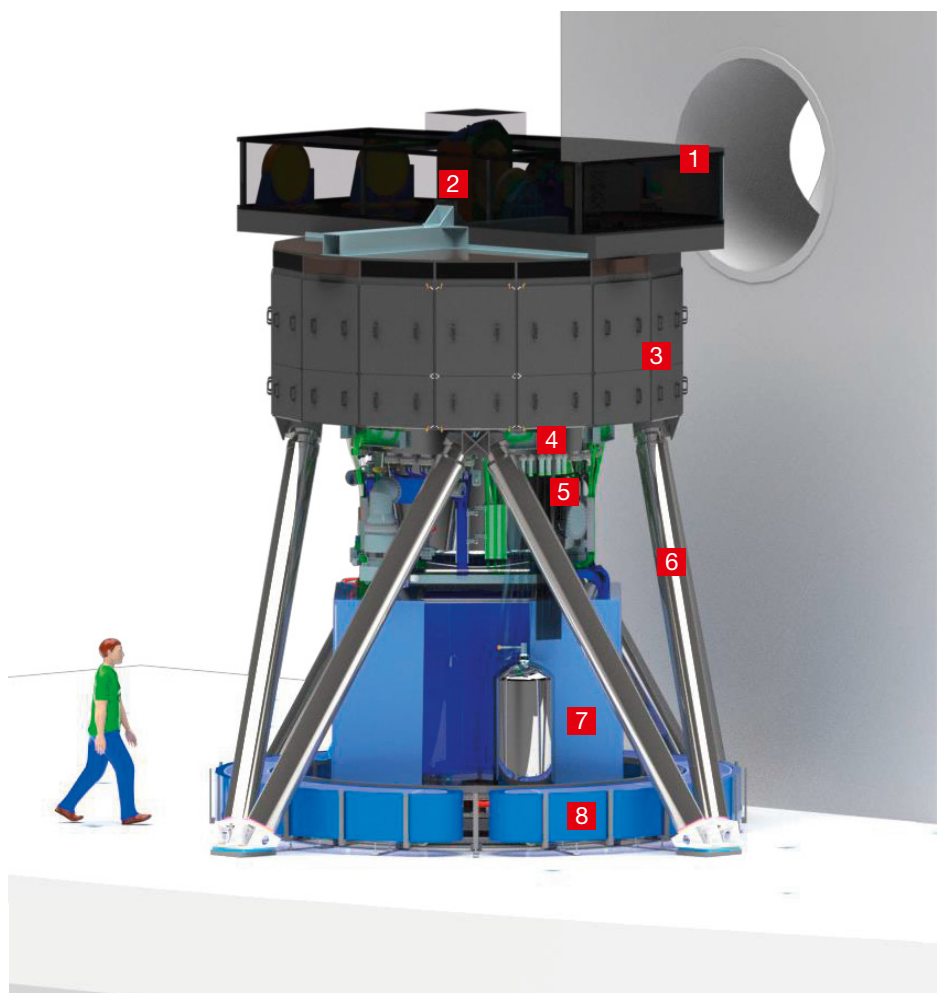

Figure 3. Rendering of MICADO in a stand-alone phase before it is integrated next to MAORY. The pre-focal station, which transfers the optical axis horizontally at a height of 6 m above the Nasmyth Platform, is indicated by the grey block. The key components of MICADO are labelled, and a person is shown for scale. For a view of MICADO in its final location next to MAORY, see the accompanying article on the MAORY system (Ciliegi et al., p. 13).

1. Calibration assembly, mounted next to relay optics, replicates ELT focal plane (later moves to MAORY bench)
2. Relay optics – transfers ELT focal plane downwards into MICADO (later exchanged for last MAORY mirror)
3. NGS WFS module, rotates under fixed cover – contains SCAO on top bench & MAORY NGS WFS on lower bench
4. Derotator
5. Cryostat, surrounded by peripheral devices
6. Support structure, for rotating mass (cryostat & NGS WFS module) as well as fixed upper platform
7. Co-rotating platform with electronics cabinets (due to cable length limitations)
8. Cable wrap for connection to external cabinets and services

coronagraphy. As an example, for HR8799e the flux from the halo of the star (type A5 with $K \sim 5.2$ magnitudes) within a 12-milliarcsecond aperture at an offset of 0.37 arcseconds is expected to be only about twice that in the PSF core of the exoplanet ($K \sim 16.5$ magnitudes). At this level, direct spectroscopy of the exoplanet is achievable within a reasonable integration time.

### Instrument design concept

As a first-light instrument, MICADO has been designed to be as simple and robust as possible. Despite this, it contains a large number of mechanisms and control systems, and requires a huge amount of software and electronics. Although one might consider these as the brain and nervous system of the instrument, we do not discuss them in detail here. Instead we focus on the opto-mechanical concept; Figure 2 provides an outline of the key modules and functions. We refer the reader to Davies et al. (2016, 2018) for technical details and additional references, with the caveat that the instrument design has evolved considerably over the last few years.

The location of the 2-m diameter cryostat, the core of the instrument, within the global architecture is illustrated in Figure 3, which shows MICADO in its stand-alone configuration. The top platform, at a height of 6 m to match the optical axis of the ELT, hosts the calibration unit which is later moved to the MAORY bench, as well as the optics that relay the ELT focal plane down into the cryostat. On the way, the optical path passes through the natural guide star wavefront sensor (NGS WFS) module which has a lower level for the MCAO low-order wavefront sensors of MAORY and an upper level containing the SCAO system. When the SCAO system is used, a dichroic is moved into the optical path to reflect out the visible light. A natural guide star can then be picked off anywhere within an offset 6 × 20 arcsecond patrol field, enabling both on-axis and off-axis correction. This versatile system makes use of a modulated pyramid wavefront sensor, providing both the very good correction needed by high-contrast imaging and partial correction on stars fainter than $V = 16$ magnitude. A calibration unit in the same volume provides the ability to measure non-common path aberrations so that they can be corrected during science observations. The NGS WFS module is mounted rigidly onto the derotator together with the cryostat, to minimise differential flexure between them while tracking. This is all held up by an octopod support structure that provides space underneath for a co-rotating platform containing much of the electronics. This both keeps some cable lengths short and reduces the volume needed by the cable-wrap that provides connections to the observatory services and to the remaining electronics.

Following a light ray entering the cryostat through the entrance window, the first mechanism it encounters contains the focal plane masks (field stops, slits, coronagraphs, calibration mask).

The aperture wheel allows those masks to be mounted closer to each other, at the same time, providing an option for rapidly blocking the light path, which is needed for mitigating persistence on the detectors. After this, the collimator, the high-resolution imager and the camera below are designed and built as complete units out of AlSi alloy, so that the coefficient of thermal expansion of the mirrors is the same as their polished NiP coating. This avoids warping of the 20–30-cm mirrors when the optics are cooled with liquid nitrogen to 80 K. The central wheel mechanism contains the two filter wheels (1.25 m in diameter, each containing 16 slots of 140-mm diameter filters), the atmospheric dispersion corrector, and the pupil wheel (where the nominal cold stop diameter is 82 mm). These are mounted into a single pre-aligned unit that can then be easily integrated into the system. The main selection mechanism, which also has a diameter of 1.3 m, enables the choice of observing mode by rotating to different positions. In the open slot, the optical path is reflected through the zoom optics, a set of fixed mirrors that comprise the high-resolution imager. When switching to the low-resolution image, the mechanism rotates in a pair of fold mirrors that bypass these.

Similarly, by inserting suitable lenses one can instead re-image the pupil for calibration. And for spectroscopy one moves in the pair of cross-dispersed gratings. After the camera, the last unit is the focal plane array in which the detectors are mounted.

## Performance

MICADO has a high throughput and a challenging wavefront error budget, both of which serve to maximise its sensitivity. The SCAO system will provide a Strehl ratio of ~ 65% in the $K$ band on-axis using a bright NGS, while the MCAO system is expected to deliver a 30–50% Strehl ratio in the $K$ band under moderate observing conditions and with a reasonable natural guide star asterism. From this we estimate the 5-hr, 5σ limiting point source magnitudes for imaging as approximately $J_{AB}$ ~ 28.6, $H_{AB}$ ~ 29.5, and $K_{AB}$ ~ 29.1 magnitudes. In doing so, we caution that a single set of performance numbers does not reflect how well one might be able to make a specific set of measurements on a realistic source, that is perhaps extended or complex. And so we emphasise that the strength of MICADO is in combining this sensitivity with a resolution 5 times better than that of the VLT. It is for this reason that we encourage potential users to try out the instrument data simulator ScopeSim[1] (recently upgraded and expanded from the original SimCADO), and to test out their own ideas about how they might use MICADO.

### Acknowledgements

The authors are grateful for the enormous effort invested into the project by the consortium and its various funding agencies. MICADO is a collaboration involving more than 100 people in Germany, France, the Netherlands, Austria, Italy, and Finland, working together with ESO.

### References

Davies, R. et al. 2016, Proc. SPIE, 9908, 99081Z
Davies, R. et al. 2018, Proc. SPIE, 10702, 107021S
Pott, J.-U. et al. 2018, Proc. SPIE, 10702, 1070290

### Links

[1] ScopeSim is available at https://scopesim.readthedocs.io/

### Notes

[a] The MICADO partners and team members can be found at http://www.mpe.mpg.de/ir/micado

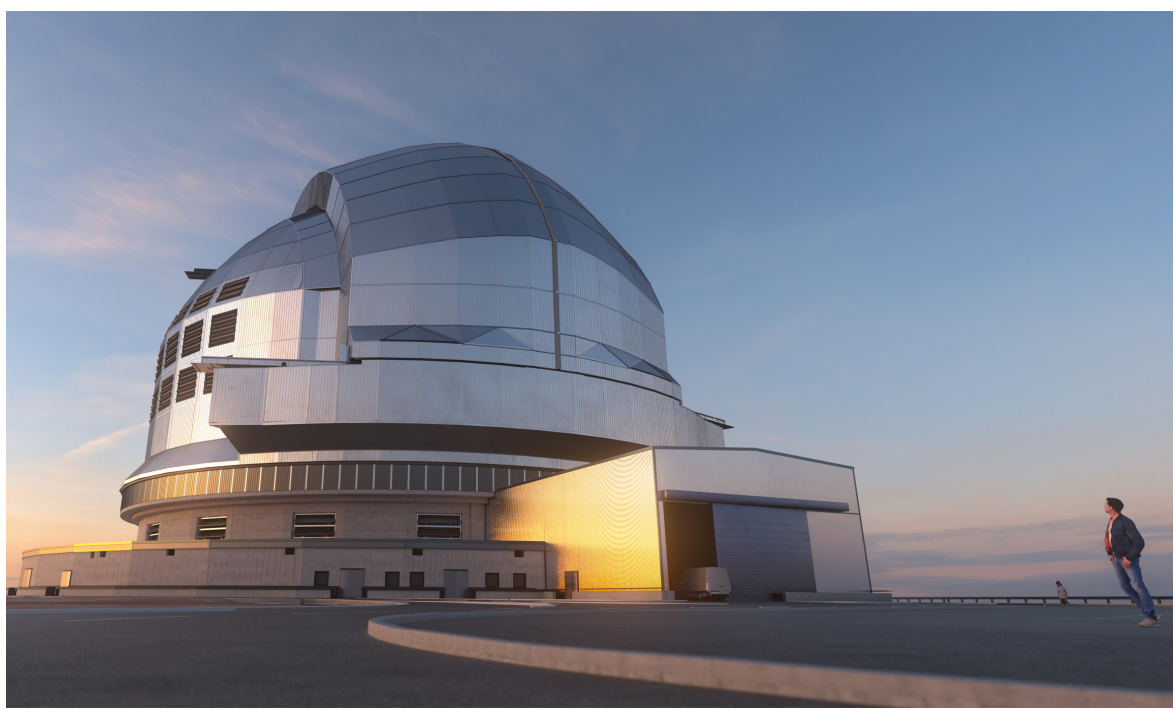

A huge structure is required to protect ESO's Extremely Large Telescope (ELT) from the elements. The telescope's structure and optical elements, including its giant 39-metre main mirror, will be housed in the largest telescope dome in the world, about 88 meters across, which is shown in this 3D rendering, along with the auxiliary building.